\documentclass[aps,prl,twocolumn,nopacs,final,letterpaper,superscriptaddress,longbibliography]{revtex4-2}

\usepackage[utf8]{inputenc}
\usepackage{calc}
\usepackage{graphicx}
\usepackage{amsmath,amssymb,amsthm}
\usepackage{mathtools}
\usepackage{txfonts}
\usepackage{bm}
\usepackage{color}
\usepackage[hidelinks]{hyperref}
\usepackage{multirow}
\graphicspath{{./figs/}}

\begin{document}

\author{Giulio Burgio}
\affiliation{Departament d'Enginyeria Inform\`atica i Matem\`atiques, Universitat Rovira i Virgili, 43007 Tarragona, Spain}
\author{Guillaume St-Onge}
\affiliation{Laboratory for the Modeling of Biological and Socio-technical Systems, Northeastern University, Boston, MA, USA}
\author{Laurent H\'{e}bert-Dufresne}
\affiliation{Vermont Complex Systems Center and Department of Computer Science, University of Vermont, Burlington, VT 05405 }

\title{Adaptive hypergraphs and the characteristic scale of higher-order contagions using generalized approximate master equations}

\begin{abstract}
People organize in groups and contagions spread across them.
A simple process, but complex to model due to dynamical correlations within groups \textit{and} between groups.
Groups can also change as agents join and leave them to avoid infection. 
To study the characteristic levels of group activity required to best model dynamics and for agents to adapt, we develop master equations for adaptive hypergraphs, finding bistability and regimes of detrimental, beneficial, and optimal rewiring, at odds with adaptation on networks. Our study paves the way for higher-order adaptation and self-organized hypergraphs.
\end{abstract}

\maketitle


Due to network effects, network neighbors are very different from random members of a population and often from each other. Any variance in heterogeneity implies that sampling a random connection is different from sampling a random node, which is the statistical bias behind the friendship paradox where ``your friends have more friends than you do.'' Accurate descriptions of dynamics on networks therefore relies on capturing important heterogeneities in how dynamical processes see a networked population. Because of their connectivity or degree, not all nodes in the network follow the same dynamics, and neither do all nodes of the same degree because of their different neighborhoods. The same is true in networks with groups. ``Your friends are parts of more groups than you are,'' but the state of group neighbors are also more correlated than expected at random. Ignoring these effects can lead to erroneous conclusions about how networks support dynamics, since degree heterogeneity \cite{st-onge2022influential} and dynamical correlations \cite{burgio2021network,burgio2023triadic} both shape critical behaviors.

To capture these effects, mathematical models often rely on \textit{approximate master equations} (AME) meant to capture local correlations and the local state of the dynamics \cite{house2008deterministic, hebert-dufresne2010propagation,marceau2010adaptive,lindquist2011effective,gleeson2013binary, o2015mathematical,unicomb2019reentrant, unicomb2021dynamics, kim2023contagion}. One can describe contagion dynamics on networks with communities or higher-order structures, captured by distinguishing nodes by their state (infectious or susceptible) and membership (how many groups they belong to) as well as groups by their size (how many nodes they contain) and composition (how many infectious nodes they contain) \cite{hebert-dufresne2010propagation}. Or, one can describe contagion dynamics on random networks by distinguishing nodes by their number of neighbors and infectious neighbors \cite{marceau2010adaptive}. These are sometimes called group and node-based AMEs respectively. The former captures the effects of groups on contagion dynamics but falls back on heterogeneous pairwise approximation~\cite{eames2002modeling} when describing random networks. The latter captures dynamical correlations with high accuracy on random networks, but cannot account for group structure.

Compared to standard heterogeneous mean-field models \cite{pastor-satorras2015epidemic,landry2020effect}, the AME frameworks stand out by tracking the full distribution of potential dynamical states within network motifs---i.e., the number of infectious nodes either within the neighborhood of a node or within a group---not just average states.
Recent results have shown that these state distributions can be very heterogeneous and even bimodal, explaining blind spots of mean-field approaches that fail to capture important dynamical regimes \cite{st-onge2021social,st-onge2021master}. Capturing these correlations is critical to describe the dynamics accurately, especially if adaptive behaviors are allowed to change the network structure in response to the composition of groups and neighborhoods.

In this paper, we combine node and group-based master equation frameworks to accurately tackle adaptive hypergraphs, where an agent can adapt its membership to groups based on their configuration.
This leads us to a series of theoretical questions: When should a group be considered \textit{active}? 
Is the definition of an active group the same for modelers trying to describe the dynamics and for agents hoping to avoid it?
Under which condition does local adaptive behavior act as a global control against contagions on hypergraphs?


\paragraph{\textbf{Generalized approximate master equations.}}

Let us consider a general binary-state dynamics on infinite-size random networks with groups.
Nodes can either be susceptible ($S$) or infected ($I$) and have a membership $m$ drawn from $g_m$. Groups are of various size $n$ drawn from $p_n$. We partition the groups and the nodes according to their local properties. Specifically, we track $C_{n,i}(t) \in [0,p_n]$, the fraction of groups of size $n$ with $i \in \lbrace 0, \dots, n \rbrace$ infected nodes at time $t$.
We also track $S_{m,l}$ and $I_{m,l} \in [0,g_m]$, the fraction of susceptible and infected nodes with membership $m$ and $l \in \lbrace 0, \dots, m \rbrace$ incident \textit{active} groups. For a given node, a group to which it belongs is active when it contains at least $\bar{i}$ infected nodes other than itself.
Accordingly, we call $l$ the active membership of a node.
These three types of compartments are also to be interpreted as joint probabilities, i.e. $C_{n,i} \equiv P(n,i)$, $S_{m,l} \equiv P(S,m,l)$, and $I_{m,l} \equiv P(I,m,l)$, with normalization $\sum_{n,i} C_{n,i} = \sum_n p_n = 1$ and $\sum_{m,l} (S_{m,l} + I_{m,l}) = \sum_m g_m = 1$.
Unless specified, sums run over all possible values.

\begin{figure}[]
    \centering
    \includegraphics[width=\linewidth]{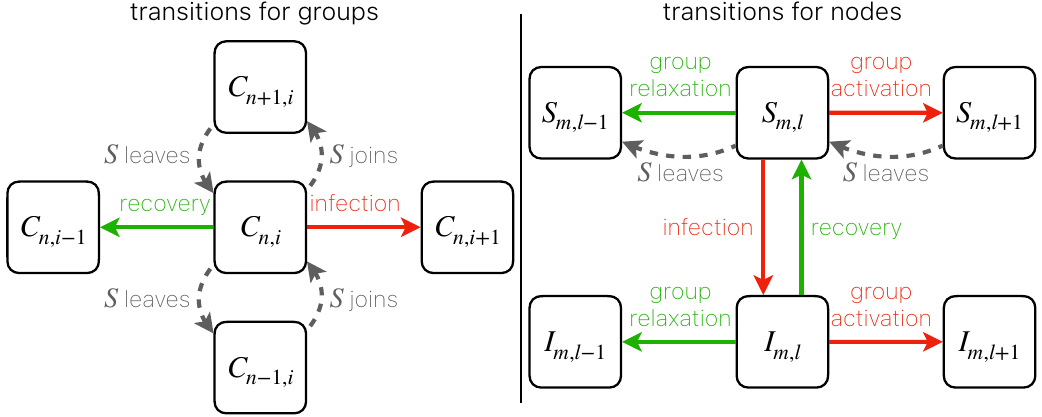}
    \caption{States and transitions for generalized approximate master equations. Groups are identified by their size $n$ and number of infectious members $i$. Nodes are identified by their group membership $m$ and the number $\ell$ of those that are considered active (i.e., containing at least $\bar{i}$ infected members). GAME transitions, Eqs.~(\ref{eq:master_equations}), correspond to binary-state ($\{S,I\}$) contagion dynamics and indicated by red and green arrows. Adaptive A-GAME transitions, Eqs.~(\ref{eq:rewiring}), are indicated by dashed arrows.}
    \label{fig:compartments}
\end{figure}

We introduce the \textit{effective} infection (recovery) rate $\bar{\beta}_{n,i}$ ($\bar{\alpha}_{n,i}$) for a node within a group of size $n$ with $i$ infected members and the effective infection (recovery) rate $\tilde{\beta}_{m,l}$ ($\tilde{\alpha}_{m,l}$) for a node of membership $m$, $l$ of which are active.
From these definitions, we introduce the following set of \textit{generalized approximate master equations} (GAME), schematized in Fig.~\ref{fig:compartments},
\begin{subequations}
\label{eq:master_equations}
\begin{align}
    \frac{\mathrm{d}C_{n,i}}{\mathrm{d}t} =& \;\bar{\alpha}_{n,i+1} (i+1) C_{n,i+1} - \bar{\alpha}_{n,i} i C_{n,i} \notag \\
                                           & + \bar{\beta}_{n,i-1} (n-i+1) C_{n,i-1} - \bar{\beta}_{n,i} (n-i) C_{n,i} \label{eq:master_equations_cni} \; \\
    \frac{\mathrm{d} S_{m,l}}{\mathrm{d} t} =& \; \tilde{\alpha}_{m,l} I_{m,l} - \tilde{\beta}_{m,l} S_{m,l} \notag \\
                                            & + \theta_S \left [ (m-l + 1) S_{m,l-1} - (m-l) S_{m,l} \right ] \label{eq:master_equations_sml} \\
                                            & + \phi_S \left [ (l + 1) S_{m,l+1} - l S_{m,l} \right ] \;, \notag \\
    \frac{\mathrm{d} I_{m,l}}{\mathrm{d} t} =& -\tilde{\alpha}_{m,l} I_{m,l} + \tilde{\beta}_{m,l} S_{m,l} \notag \\
                                            & + \theta_I \left [ (m-l + 1) I_{m,l-1} - (m-l) I_{m,l} \right ] \label{eq:master_equations_iml} \\
                                            & + \phi_I \left [ (l + 1) I_{m,l+1} - l I_{m,l} \right ] \;. \notag
\end{align}
\end{subequations}
The four mean fields are calculated as
\begin{subequations}
\label{eq:mf}
\begin{align}
    \theta_S &= \frac{\sum_{n} (n-\bar{i}+1) (n-\bar{i}) C_{n,\bar{i}-1} \bar{\beta}_{n,\bar{i}-1}}{\sum_{n,i \leq \bar{i}-1} (n-i) C_{n,i}} \;, \label{eq:mf_th_s}\\
    \phi_S &= \frac{\sum_{n} (n-\bar{i}) \bar{i} C_{n,\bar{i}} \bar{\alpha}_{n,\bar{i}}}{\sum_{n,i>\bar{i}-1} (n-i) C_{n,i}} \;, \label{eq:mf_ph_s}\\
    \theta_I &= \frac{\sum_{n} \bar{i} (n-\bar{i}) C_{n,\bar{i}} \bar{\beta}_{n,\bar{i}}}{\sum_{n,i \leq \bar{i}} i C_{n,i}} \;, \label{eq:mf_th_i}\\
    \phi_I &= \frac{\sum_{n} (\bar{i}+1) \bar{i} C_{n,\bar{i}+1} \bar{\alpha}_{n,\bar{i}+1}}{\sum_{n,i>\bar{i}} i C_{n,i}} \;. \label{eq:mf_ph_i}
\end{align}
\end{subequations}
These quantities are the average rates at which inactive groups become active ($\theta$) or vice versa ($\phi$) given that we know the state of one of their members (subscript). We therefore sum over all groups eligible for the transition (e.g. with one too few or too many infectious nodes) and count the number of nodes therein whose state matches that of the node of interest. This gives us a biased distribution over states, renormalized with the sum in the denominator, and over which we average the local rate of transition (every factor in the numerator not in the denominator).

To close the GAME, we need to estimate the previously introduced effective transition rates, $\bar{\alpha}_{n,i}$, $\bar{\beta}_{n,i}$, $\tilde{\alpha}_{m,l}$ and $\tilde{\beta}_{m,l}$. 
We can calculate these rates with mean-field arguments, but the form of this calculation depends on the nature of the dynamics. Specifically, we use two different approaches based on whether the dynamics operate at the node level (i.e., a node gets infected based on its total number of infectious neighbors) or at the group level (i.e., a node gets infected through each group based on the number of infectious members therein). We show the simplest version here and derive the other in the Supplemental Material.


\paragraph{\textbf{Group-based dynamics.}}

We consider general continuous-time Markov processes where groups are the main actors responsible for infections.
Specifically, a susceptible node in a group of size $n$ with $i$ infectious receives an infection rate of $\lambda(n,i)$ from this group.
The effective rate $\bar{\beta}_{n,i}$ thus becomes $\lambda(n,i) + \bar{\lambda}_i$, where $\bar{\lambda}_i$ is the infection rate from all the external groups, reading
\begin{align}
    \bar{\lambda}_i = \begin{dcases}
                        \frac{\sum_{m,l} (m-l) S_{m,l} \left[ (m\!-\!l\!-\! 1) \bar{\lambda}_{i<\bar{i}} + l \bar{\lambda}_{i \geq \bar{i}} \right]}{\sum_{m,l} (m-l) S_{m,l} } & \text{if } i < \bar{i}\;,
                        \\
                        \frac{\sum_{m,l} l S_{m,l} \left[ (m-l) \bar{\lambda}_{i<\bar{i}} + (l-1) \bar{\lambda}_{i \geq \bar{i}} \right]}{\sum_{m,l} l S_{m,l} } & \text{if } i \geq \bar{i}\;,
                    \end{dcases}
    \label{eq:lambda_i}
\end{align}
Where $\bar{\lambda}_{i \geq \bar{i}}$ and $\bar{\lambda}_{i < \bar{i}}$ are the mean infection rates of a random external inactive or active group respectively---they do not depend on $i$. They read as:
\begin{align}
    \bar{\lambda}_{i < \bar{i}} &= \frac{\sum_{n,i < \bar{i}} (n-i) C_{n,i} \lambda(n,i)}{\sum_{n,i < \bar{i}} (n-i) C_{n,i}} 
    \;, \label{eq:lambda_ilj} \\
    \bar{\lambda}_{i \geq \bar{i}} &= \frac{\sum_{n,i \geq \bar{i}} (n-i) C_{n,i} \lambda(n,i)}{\sum_{n,i \geq \bar{i}} (n-i) C_{n,i}} 
    \label{eq:lambda_igj} \;.
\end{align}
Similarly, the effective rate $\tilde{\beta}_{m,l}$ simply reads
\begin{equation}
     \tilde{\beta}_{m,l} = (m-l) \bar{\lambda}_{i < \bar{i}} + l \bar{\lambda}_{i \geq \bar{i}}  \;.
     \label{eq:group_beta_ml} 
\end{equation}

Finally, the rates $\bar{\alpha}_{n,i}$ and $\tilde{\alpha}_{m,l}$ can be computed analogously (see Supplemental Material). For the contagion dynamics we analyze in the following, recovery is a spontaneous node transition, for which we use a simple constant rate $\bar{\alpha}_{n,i} = \tilde{\alpha}_{m,l} = 1$.

\begin{figure*}[]
    \centering
    \includegraphics[width=0.85\linewidth]{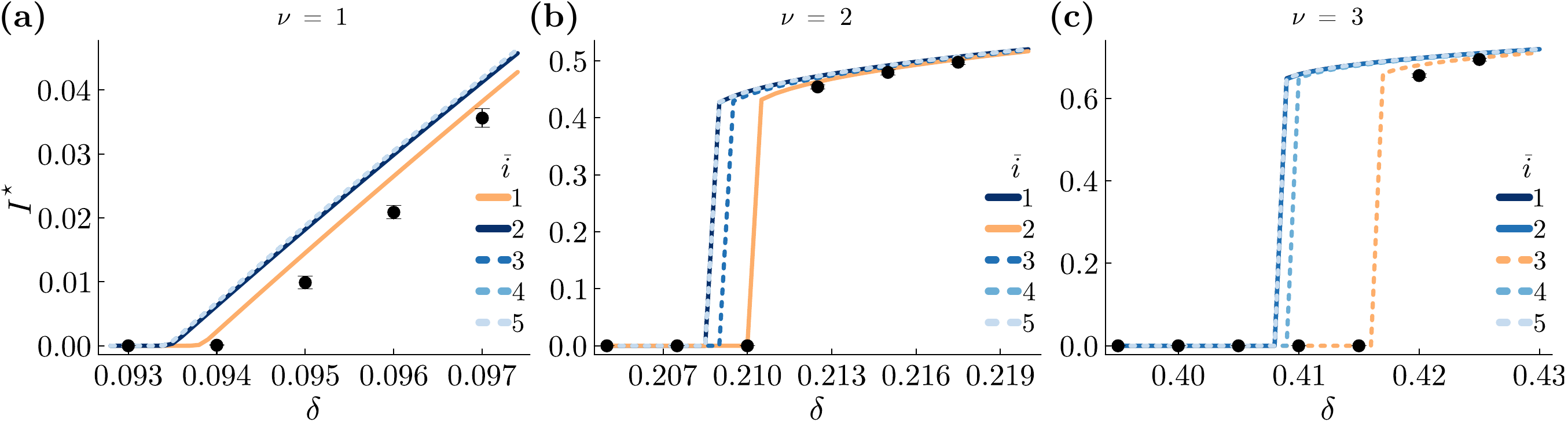}
    \caption{Equilibrium prevalence, $I^\star$, obtained on random 3-regular 5-uniform hypergraphs under group-based dynamics (Eqs.~(\ref{eq:lambda_i})-(\ref{eq:group_beta_ml})) considering a threshold-linear infection rate $\lambda(n,i) = \delta i$ if $i\geq\nu$ and $0$ otherwise, for (a) $\nu=1$ (initial prevalence, $I(0)=0.05$), (b) $\nu=2$ ($I(0)=0.8$) and (c) $\nu=3$ ($I(0)=0.8$). Solid and dashed lines represent the results obtained integrating Eqs.~(\ref{eq:master_equations}) for $\bar{i}\in\{1,\dots,5\}$, while points and error bars (when visible) denote averages and standard errors over $20$ random realizations resulting from Monte Carlo simulations performed on hypergraphs with $N=5\times 10^4$ nodes. The most accurate model is always the one having $\bar{i}=\nu$ (orange).}
    \label{fig:threshold}
\end{figure*}

\paragraph{\textbf{Results on static networks.}}

We illustrate the high accuracy of the GAME against simulations on static hypergraphs in Fig.~\ref{fig:threshold}. More importantly, this allows us to show which characteristic activity level $\bar{i}$ best captures the dynamics. We test diverse contagions functions, using $\lambda(n,i) = \delta i$ when $i\geq \nu$ and 0 otherwise. This transmission function produces classic simple contagion when $\nu = 1$ and so-called threshold or complex contagion~\cite{PhysRevLett.92.218701, lehmann2018complex,guilbeault2018complex} where a minimum number of infectious neighbors is required for transmission when $\nu > 1$.

We find that against both simple and complex contagions, the optimal description is the one whose characteristic activity level $\bar{i}$ matches the minimum number of infectious neighbors required for transmission, $\nu$. One could expect that a higher $\bar{i}$ might be useful to distinguish large groups that are actively transmitting from small groups where nodes were infected through some other groups. Yet, the most straightforward answer is the best: The optimal model is the one that tags as \textit{active} only the groups that \textit{can} transmit.

\paragraph{\textbf{The A-GAME: Adaptive hypergraphs.}}

We now consider the case where susceptible nodes can rewire away from active groups. This higher-order adaptation mechanism is a hypergraph generalization of the adaptive network model by Gross et al. \cite{gross2006epidemic}.
To quantify the information nodes have about groups before joining them, we define the probability $\eta$ that rewiring is targeted towards inactive groups, as opposed to random groups in any state (notice that the model in Gross et al. \cite{gross2006epidemic} only describes the case $\eta=1$).
With group rewiring, the average group size and average group membership are conserved, but the active membership of nodes and the size distribution of groups (and consequently the degree distribution of the projected network) are allowed to change as an adaptive response to the contagion. In theory, this model can interpolate between a complete network with a giant infinite group and a sparse regular network.

Rewiring adds the following transitions to our system,
\begin{subequations}
\label{eq:rewiring}
\begin{align}
    \frac{\mathrm{d}C^{\textrm{a}}_{n,i}}{\mathrm{d}t} =& \; \gamma {\bf 1}_{i \geq \bar{i}} \left[(n+1-i)C_{n+1,i} - (n-i)C_{n,i}\right] \label{eq:rewiring_cni} \\
            &+ \left( \frac{\eta {\bf 1}_{i < \bar{i}}}{C_{i<\bar{i}}} + 1-\eta \right) \times \left[\gamma \Omega_{S\vert i\geq \bar{i}}C_{n-1,i} - \gamma \Omega_{S\vert i\geq \bar{i}} C_{n,i} \right] \notag \;, \\
    \frac{\mathrm{d} S^{\textrm{a}}_{m,l}}{\mathrm{d} t} =& \; \gamma \left[\eta + (1-\eta) C_{i<\bar{i}}\right] \left[(l+1)S_{m,l+1}-lS_{m,l}\right] \;. \label{eq:rewiring_sml}
\end{align}
\end{subequations}
These transitions are tagged with `${\textrm{a}}$' for \textit{adaptive} and  added to Eqs.~(\ref{eq:master_equations_cni})-(\ref{eq:master_equations_sml}).
The terms in Eq.~(\ref{eq:rewiring_cni}) accounts for susceptible nodes leaving and joining groups, respectively, while Eq.~(\ref{eq:rewiring_sml}) only needs to account for susceptible nodes rewiring away from their active groups. To calculate these rates, we define ${\bf 1}_\text{p}$, which equals $1$ if proposition p is true and $0$ otherwise, to tag active groups whose members might escape and inactive groups which might be targeted; and also $C_{i<\bar{i}} = \sum_{n,i<\bar{i}} C_{n,i}$ and  $\Omega_{S \vert i \geq \bar{i}} = \sum_{n,i\geq \bar{i}} (n-i) C_{n,i}$. Note that we formulate a more general adaptive hypergraph, where both susceptible and infectious agents can rewire \cite{scarpino2016effect}, in our Supplemental Material.

\begin{figure*}[]
    \centering
    \includegraphics[width=0.85\linewidth]{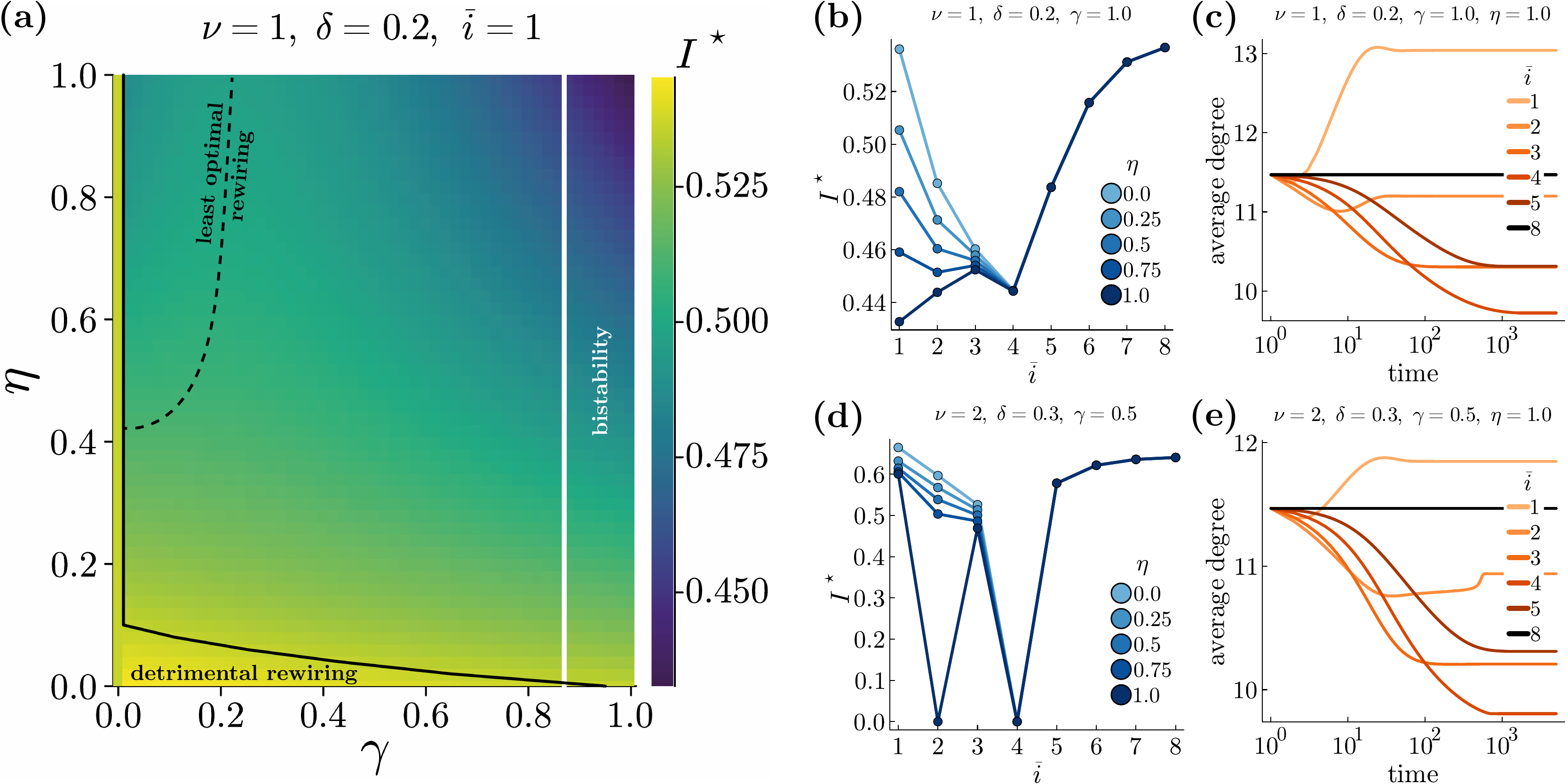}
    \caption{Results for adaptive hypergraphs (Eqs.~(\ref{eq:rewiring})). The initial size distribution is a truncated Poisson with mode $n=4$ and support $\{2,\dots,8\}$ ($n=8$ being a maximal group size), yielding an average group size $\langle n\rangle = 4.196$. All nodes have membership $m=3$.
    (a) Phase diagram for a simple contagion ($\nu = 1$) showing the equilibrium prevalence $I^\star$ against rewiring rate $\gamma$ and rewiring accuracy $\eta$, when strategy $\bar{i}=\nu=1$ is used. We find a bistable region when rewiring is fast enough (notice that the invasion threshold is independent from $\eta$); a low-accuracy region where rewiring is detrimental (on the left of the black line), a slow-rewiring region where, for high enough accuracy, prevalence first increases and then decreases with $\gamma>0$, defining a local prevalence maximum or least optimal rewiring rate (black dashed line).
    (b) Varying $\bar{i}$ we find two main strategies for the agents to reduce (and possibly eradicate) contagions. At high $\gamma$ and $\eta$, it is optimal for agents to target the dynamics by setting $\bar{i}=\nu$, which is also how modelers can optimize their model as seen in Fig.~\ref{fig:threshold}. At low $\eta$ or low $\gamma$ (see Supplemental Material for the latter case), it is optimal to target instead the structure and minimize degree by setting $\bar{i} = 4 \approx \langle n\rangle$.
    (c) Time evolution of the average degree in case (b) for $\eta = 1.0$. Strategies $\bar{i}=\nu$ and $\bar{i} = 4 \approx \langle n\rangle$ are the two best, but only the latter explicitly targets the structure in order to reduce the number of contacts. (d \& e) Analogous to panels (b \& c) but for complex contagion ($\nu = 2$). 
    }
    \label{fig:adaptive}
\end{figure*}

\paragraph{\textbf{Results on adaptive networks.}} We experiment with adaptive hypergraphs on both simple ($\nu=1$) and complex contagions ($\nu=2$) in Fig.~\ref{fig:adaptive}. The phase diagram of the dynamics highlights a few important results. There exists a bistable region for any accuracy when rewiring is fast enough, generalizing the results from Gross \textit{et al}.~\cite{gross2006epidemic}. But more importantly, the higher-order organization leads to two previously unseen phenomena. First, a low-accuracy region of detrimental rewiring, where prevalence is higher than with no rewiring. Second, a slow-rewiring region where, for high enough accuracy, prevalence is lower than with no rewiring, but increases with $\gamma$ before eventually decreasing, defining a value of least optimal rewiring rate for each $\eta$.

To better understand the rich phenomenology of higher-order adaptation, we identify two strategies for agents in the network to control the dynamics:
\begin{enumerate}
    \item Nodes can avoid the contagion. This is optimal when targeting is both fast (high $\gamma$) and accurate (high $\eta$). To do so, nodes have to mimic modelers and set $\bar{i}=\nu$. Being enough reactive and precise, they can manage to escape infection without necessarily lower the connectivity of the structure; 
    \item Nodes can avoid large groups. This is optimal when rewiring is slow (low $\gamma$) or targeting 
 is poor (low $\eta$). By setting $\bar{i} \approx \langle n \rangle$, the average group size, nodes rewire away from groups that are larger than average, thereby minimizing their average degree and the probability of getting infected.
\end{enumerate}

Importantly, to say whether the rewiring is `fast' or `slow' and `accurate' or `poor', we have to compare the system with the state of its static counterpart. There is therefore no clear boundary defining which strategy is optimal. 
In fact, in intermediate regimes, both strategies can work just as well but an intermediary strategy ($\nu < \bar{i} < \langle n \rangle$) may not, see Fig.~\ref{fig:adaptive}(d). We hypothesize that this is because the mechanisms underlying these strategies are actually in opposition.

In Fig.~\ref{fig:adaptive}(c \& e), we show the average degree of the adaptive hypergraphs under different rewiring strategies. We see that the $\bar{i}=\nu$ strategy works despite slightly decreasing ($\nu=2$) or even increasing ($\nu=1$) the connectivity of the system.
As expected, the $\bar{i}\approx\langle n \rangle$ strategy works by having susceptible agents avoid groups larger than average and therefore create a more uniform sparse network, actually minimizing the average degree. In accordance to our hypothesis, the two adaptive strategies work in different ways, targeting either dynamics or structure.

A similar logic explains the observed region of least-optimal rewiring rates. Larger groups reach $i = \bar{i}$ faster on average and slow rewiring allows $i$ to significantly correlate with group size before the typical rewiring time ($1/\gamma$). For high enough accuracy, susceptible agents then preferentially migrate to smaller groups, decreasing the average degree (see Supplemental Material). Conversely, fast rewiring makes targeting the dynamics the optimal strategy. In between, we find a least optimal rewiring rate that is too slow to avoid the dynamics but too fast to minimize the degree.

\paragraph{\textbf{Discussion.}}

We studied contagions on static and adaptive hypergraphs by developing a very general model to capture dynamical correlations both within groups and across groups.
To do so, we introduced the notion of characteristic scale $\bar{i}$ of a contagion to tag groups as active ($i\geq\bar{i}$) or inactive ($i<\bar{i}$) based on the number of infectious nodes $i$ they contain. Our GAME thus generalizes both node-based and group-based AME frameworks, which are respectively recovered when one collapses groups to pairwise edges or considers all groups as equivalently active. Hence, whether the aim is to describe binary-state dynamics on networks or hypergraphs, on static or adaptive structures, it's in the GAME.

We then asked three questions. What is the characteristic scale $\bar{i}$ such that our mathematics best fit simulations on static hypergraphs? What is the characteristic scale $\bar{i}$ such that agents can best avoid the contagion? And how similar are those two answers?

We found that modelers should always use a characteristic scale $\bar{i}=\nu$ set by the number of infectious neighbor $\nu$ necessary for infection. However, agents have more options if they want to avoid the contagion. When rewiring is fast and accurate, agents can act as modelers and set $\bar{i}=\nu$ to minimize the contagion events without minimizing their connectivity. When rewiring is slow or inaccurate, agents should instead aim to minimize their degree by rewiring away from large groups based on $\bar{i}\approx\langle n \rangle$, the average group size.

Altogether, this work introduced adaptive hypergraphs, which are not as constrained as most adaptive network models for their density or average degree is not fixed over time, which enables them to organize in diverse ways. Yet, the conserved quantities of average hyperdegree and hyperedge size allowed us to formulate a very accurate and general model based on approximate master equations. We hope that our contributions will inspire future work on self-organized hypergraph and group structures. \\

\paragraph{\textbf{Acknowledgements.}} The authors thank Antoine Allard and Alex Arenas for discussions and feedback. G.B.\ acknowledges financial support from the European Union's Horizon 2020 research and innovation program under the Marie Sk\l{}odowska-Curie Grant Agreement No.\ 945413 and from the Universitat Rovira i Virgili (URV), G.S. from the Fonds de recherche du Québec -- Nature et technologies (project 313475), and L.H.-D. from the National Institutes of Health 1P20 GM125498-01 Centers of Biomedical Research Excellence Award. 

\appendix

\end{document}


\author{Giulio Burgio}
\affiliation{Departament d'Enginyeria Inform\`atica i Matem\`atiques, Universitat Rovira i Virgili, 43007 Tarragona, Spain}
\author{Guillaume St-Onge}
\affiliation{Laboratory for the Modeling of Biological and Socio-technical Systems, Northeastern University, Boston, MA, USA}
\author{Laurent H\'{e}bert-Dufresne}
\affiliation{Vermont Complex Systems Center and Department of Computer Science, University of Vermont, Burlington, VT 05405 }

\title{Adaptive hypergraphs and the characteristic scale of higher-order contagions using generalized approximate master equations\\ \vspace{1ex}\normalsize--- Supplemental Material ---}

\maketitle

\tableofcontents

\section{Generalized Approximate Master Equations (GAME)}

Recall that the GAME on static networks involves the following transitions

\begin{subequations}
\label{eq:master_equations}
\begin{align}
    \frac{\mathrm{d}C_{n,i}}{\mathrm{d}t} =& \;\bar{\alpha}_{n,i+1} (i+1) C_{n,i+1} - \bar{\alpha}_{n,i} i C_{n,i}  + \bar{\beta}_{n,i-1} (n-i+1) C_{n,i-1} - \bar{\beta}_{n,i} (n-i) C_{n,i} \label{eq:master_equations_cni} \; \\
    \frac{\mathrm{d} S_{m,l}}{\mathrm{d} t} =& \; \tilde{\alpha}_{m,l} I_{m,l} - \tilde{\beta}_{m,l} S_{m,l}  + \theta_S \left [ (m-l + 1) S_{m,l-1} - (m-l) S_{m,l} \right ]  + \phi_S \left [ (l + 1) S_{m,l+1} - l S_{m,l} \right ] \;, \label{eq:master_equations_sml} \\
    \frac{\mathrm{d} I_{m,l}}{\mathrm{d} t} =& -\tilde{\alpha}_{m,l} I_{m,l} + \tilde{\beta}_{m,l} S_{m,l} + \theta_I \left [ (m-l + 1) I_{m,l-1} - (m-l) I_{m,l} \right ] + \phi_I \left [ (l + 1) I_{m,l+1} - l I_{m,l} \right ] \;, \label{eq:master_equations_iml}
\end{align}
\end{subequations}
where the four mean fields are calculated as
\begin{subequations}
\label{eq:mf}
\begin{align}
    \theta_S &= \frac{\sum_{n} (n-\bar{i}+1) (n-\bar{i}) C_{n,\bar{i}-1} \bar{\beta}_{n,\bar{i}-1}}{\sum_{n,i \leq \bar{i}-1} (n-i) C_{n,i}} \;, \label{eq:mf_th_s}\\
    \phi_S &= \frac{\sum_{n} (n-\bar{i}) \bar{i} C_{n,\bar{i}} \bar{\alpha}_{n,\bar{i}}}{\sum_{n,i>\bar{i}-1} (n-i) C_{n,i}} \;, \label{eq:mf_ph_s}\\
    \theta_I &= \frac{\sum_{n} \bar{i} (n-\bar{i}) C_{n,\bar{i}} \bar{\beta}_{n,\bar{i}}}{\sum_{n,i \leq \bar{i}} i C_{n,i}} \;, \label{eq:mf_th_i}\\
    \phi_I &= \frac{\sum_{n} (\bar{i}+1) \bar{i} C_{n,\bar{i}+1} \bar{\alpha}_{n,\bar{i}+1}}{\sum_{n,i>\bar{i}} i C_{n,i}} \;. \label{eq:mf_ph_i}
\end{align}
\end{subequations}

\section{Adaptive Generalized Approximate Master Equations (A-GAME)}

For a general adaptive hypergraphs model, we allow both susceptible and infectious nodes to rewire away from groups, respectively at rates $\gamma_S$ and $\gamma_I$. This leads to the following additional adaptive terms (tagged with the `$\mathrm{a}$' superscript) to the static GAME framework
\begin{subequations}
\label{eq:rewiring}
\begin{align}
    \frac{\mathrm{d}C^\mathrm{a}_{n,i}}{\mathrm{d}t} =& \; \gamma_S {\bf 1}_{i \geq \bar{i}} \left[(n+1-i)C_{n+1,i} - (n-i)C_{n,i}\right] + \gamma_I \left[{\bf 1}_{i \geq \bar{i}} (i+1)C_{n+1,i+1} - {\bf 1}_{i > \bar{i}} iC_{n,i}\right] \notag \\
            &+ \left( \frac{\eta {\bf 1}_{i < \bar{i}}}{C_{i<\bar{i}}} + 1-\eta \right) \left[\gamma_S \Omega_{S\vert i\geq \bar{i}}C_{n-1,i} - \left( \gamma_S \Omega_{S\vert i\geq \bar{i}} + \gamma_I \Omega_{I\vert i>\bar{i}}\right)C_{n,i} \right] + \left( \frac{\eta {\bf 1}_{i \leq \bar{i}}}{C_{i<\bar{i}}} + 1-\eta \right) \gamma_I \Omega_{I\vert i>\bar{i}} C_{n-1,i-1} \label{eq:rewiring_cni} \;,
           \\ 
    \frac{\mathrm{d} S^\mathrm{a}_{m,l}}{\mathrm{d} t} =& \; \gamma_S \left[\eta + (1-\eta) C_{i<\bar{i}}\right] \left[(l+1)S_{m,l+1}-lS_{m,l}\right] \notag \\
            &+ \gamma_I \Omega_{I\vert i>\bar{i}} \left( \frac{\eta}{\Omega_{S\vert i<\bar{i}}} + \frac{1-\eta}{\Omega_{S}} \right) \Omega_{S\vert i=\bar{i}-1} \left[(m-l+1)S_{m,l-1}-(m-l)S_{m,l}\right] \;, \label{eq:rewiring_sml} 
            \\
    \frac{\mathrm{d} I^\mathrm{a}_{m,l}}{\mathrm{d} t} =& \; \gamma_I \left[\eta + (1-\eta) C_{i<\bar{i}}\right] \left[(l+1)I_{m,l+1}-lI_{m,l}\right] \notag \\
            &+ \gamma_I ~\frac{\Gamma_{I\vert i=\bar{i}+1}}{\Omega_{I\vert i>\bar{i}}} \left[(l+1)I_{m,l+1}-lI_{m,l}\right] + \gamma_I \Omega_{I\vert i>\bar{i}} (1-\eta) \frac{\Omega_{I\vert i=\bar{i}}}{\Omega_{I}} \left[(m-l+1)I_{m,l-1}-(m-l)I_{m,l}\right] \;, \label{eq:rewiring_iml}
\end{align}
\end{subequations}
These terms are simply added to the previous system of equations. Here ${\bf 1}_\text{p}$ equals $1$ if proposition p is true and $0$ otherwise. We defined $C_{i<\bar{i}} = \sum_{n,i<\bar{i}} C_{n,i}$, $\Omega_S = \sum_{n,i} (n-i) C_{n,i}$, $\Omega_{S \vert i \geq \bar{i}} = \sum_{n,i\geq \bar{i}} (n-i) C_{n,i}$, $\Omega_{S \vert i < \bar{i}} = \sum_{n,i < \bar{i}} (n-i) C_{n,i}$, $\Omega_{S\vert i = \bar{i}-1} = \sum_{n} (n-\bar{i}+1) C_{n,\bar{i}-1}$, $\Omega_I = \sum_{n,i} i C_{n,i}$, $\Omega_{I\vert i > \bar{i}} = \sum_{n,i > \bar{i}} i C_{n,i}$, $\Omega_{I\vert i = \bar{i}} = \sum_{n} \bar{i} C_{n,\bar{i}}$, and $\Gamma_{I\vert i = \bar{i}+1} = \sum_{n} \bar{i}(\bar{i}+1) C_{n,\bar{i}+1}$. The first row in Eq.~(\ref{eq:rewiring_cni}) account for nodes leaving groups, the second for nodes joining groups. The first term in Eq.~(\ref{eq:rewiring_sml}) accounts for S nodes rewiring away, while the second term for I nodes rewiring to groups with $i=\bar{i}-1$, making them active for the S nodes therein. Lastly, the first term in Eq.~(\ref{eq:rewiring_iml}) accounts for I nodes rewiring away, while the second (third) term accounts for I nodes leaving (joining) groups with $i=\bar{i}+1$ ($i=\bar{i}$), making them inactive (active) for the I nodes therein. The simpler version where only S nodes rewire is recovered by taking $\gamma_I=0$.

\section{Node-based dynamics}

We consider general continuous-time Markov processes where infected nodes transition to the susceptible state at rate $\alpha(k,\ell)$, where $k$ is the degree of the node and $\ell \in \lbrace 0, \dots, k \rbrace$ is its infected degree.
Similarly, susceptible nodes transition to the infected state at rate $\beta(k,\ell)$. 
Importantly, degree $k$ and infected degree $\ell$ are total quantities of nodes summed over one or many groups. 
The dynamics therefore ignores how these quantities are distributed over groups.
Within this general node-based process, we can calculate the mean-field transition rates as follows.

Let us consider the first effective rate $\bar{\alpha}_{n,i}$. If we pick an infected node in clique of size $n$, of which $i$ are infected, the degree of this node can be decomposed as $k = n - 1 + r$, where $r$ is the \textit{excess} degree, due to memberships to other cliques. Similarly, the infected degree can be decomposed as $\ell = i-1 + s$, where $s$ is the excess infected degree.
If $r$ and $s$ where specified, then the recovery rate of this infected node would simply be $\alpha(n-1+r,i-1+s)$.
In this version of the GAME, we approximate $\bar{\alpha}_{n,i}$ by averaging over a joint distribution
\begin{align*}
    \bar{\alpha}_{n,i} = \sum_{r,s} \alpha(n-1+r,i-1+s) P(r,s|n,i,I) \;,
\end{align*}
where the distribution $P(r,s|n,i,I)$ is to be determined.
It is the probability that an infected node in a clique of size $n$ with $i$ infected nodes has an excess degree of $r$ and infected excess degree of $s$.
For this task, we leverage the properties of probability generating functions (PGFs).

If we take a random \textit{external} clique to which this infected node belongs, the contribution to its degree and infected degree is associated with two PGFs, depending whether or not the clique is active for it.
If it is not (i.e., the clique contains at most $\bar{i}$ infected nodes, including the focal node), then the appropriate PGF to use is
\begin{align}
    K_I^{i \leq \bar{i}}(x,y) &= \frac{\sum_{n,i\leq \bar{i}} i C_{n,i} x^{n-1} y^{i-1}}{\sum_{n,i \leq \bar{i}} i C_{n,i}} \;,
\end{align}
otherwise, it is
\begin{align}
    K_I^{i > \bar{i}}(x,y) &= \frac{\sum_{n,i>\bar{i}} i C_{n,i} x^{n-1}y^{i-1}}{\sum_{n,i>\bar{i}} i C_{n,i}} \;.
\end{align}

The excess membership and the excess active membership of the node are also unspecified but are described by the following PGF
\begin{align}
    G_{I}^i(x,y) = \begin{dcases}
                \frac{\sum_{m,l} (m-l) I_{m,l} x^{m-l-1} y^{l}}{\sum_{m,l} (m-l) I_{m,l}} & \text{if } i \leq \bar{i}\;, \\
                \frac{\sum_{m,l} l I_{m,l} x^{m-l} y^{l-1}}{\sum_{m,l} l I_{m,l}} & \text{if } i > \bar{i}\;.
             \end{dcases}
\end{align}
Assuming that the contribution to $r$ and $s$ from each clique is independent, the bivariate PGF characterizing the total excess degree and excess infected degree is
\begin{align}
    E_I^i(x,y) = G_I^i\left( K_I^{i \leq \bar{i}}(x,y), K_I^{i>\bar{i}}(x,y) \right) \;.
\end{align}
More specifically, we have
\begin{align*}
    E_I^i(x,y) = \sum_{r,s} P(r,s|n,i,I) x^r y^s \;.
\end{align*}
To highlight that the expected values are obtained with the distribution associated with the above PGF, we write
\begin{align*}
    \bar{\alpha}_{n,i} \equiv \langle \alpha(n-1+r,i-1+s) \rangle_{E_I^i} \;.
\end{align*}
Very similar steps are followed for the effective infection rate $\bar{\beta}_{n,i}$, leading to
\begin{align}
    K_S^{i<\bar{i}}(x,y) &= \frac{\sum_{n,i<\bar{i}} (n-i) C_{n,i} x^{n-1} y^{i}}{\sum_{n,i<\bar{i}} (n-i) C_{n,i}} \;, \\
    K_S^{i \geq \bar{i}}(x,y) &= \frac{\sum_{n,i \geq \bar{i}} (n-i) C_{n,i} x^{n-1}y^{i}}{\sum_{n,i \geq \bar{i}} (n-i) C_{n,i}} \;, \\
    G_S^i(x,y) &= \begin{dcases}
                \frac{\sum_{m,l} (m-l) S_{m,l} x^{m-l-1} y^{l}}{\sum_{m,l} (m-l) S_{m,l}} & \text{if } i < \bar{i}\;, \\
                \frac{\sum_{m,l} l S_{m,l} x^{m-l} y^{l-1}}{\sum_{m,l} l S_{m,l}} & \text{if } i \geq \bar{i}\;,
             \end{dcases} \\
    E_S^i(x,y) &= G_S^i\left( K_S^{i<\bar{i}}(x,y), K_S^{i \geq \bar{i}}(x,y) \right) \;,
\end{align}
and eventually to
\begin{align*}
    \bar{\beta}_{n,i} \equiv \langle \beta(n-1+r,i+s) \rangle_{E_S^i} \;.
\end{align*}

The construction of the PGFs to compute $\tilde{\alpha}_{m,l}$ and $\tilde{\beta}_{m,l}$ is actually simpler.
For an infected node with membership $m$ and $l$ of them active, the PGF for its degree $k$ and infected degree $\ell$ is
\begin{align}
    E_I^{m,l}(x,y) = \left [ K_I^{i \leq \bar{i}}(x,y)\right ]^{m-l} \left [ K_I^{i>\bar{i}}(x,y)\right ]^l \;,
\end{align}
which means that
\begin{align*}
    E_I^{m,l}(x,y) = \sum_{k,\ell} P(k,\ell|m,l,I) x^k y^\ell \;.
\end{align*}
The effective recovery rate is then
\begin{align*}
    \tilde{\alpha}_{m,l} \equiv \langle \alpha(k,\ell) \rangle_{E_I^{m,l}} \;.
\end{align*}
Analogously, for a susceptible node, we have
\begin{align}
    E_S^{m,l}(x,y) &= \left [ K_S^{i<\bar{i}}(x,y)\right ]^{m-l} \left [ K_S^{i \geq \bar{i}}(x,y)\right ]^l \;,
\end{align}
and the effective infection rate reads
\begin{align*}
    \tilde{\beta}_{m,l} \equiv \langle \beta(k,\ell) \rangle_{E_S^{m,l}} \;.
\end{align*}

This closes the GAME and A-GAME for node-based dynamics. The group-based dynamics common in the literature on higher-order network is simpler and covered in the main text.

\section{Computing the effective transition rates}

For the node-based dynamics, the effective transition rates $\bar{\alpha}_{n,i}$, $\bar{\beta}_{n,i}$, $\tilde{\alpha}_{m,l}$ and $\tilde{\beta}_{m,l}$ require that we extract the joint distributions from the bivariate PGFs $E_I^i$, $E_S^i$, $E_I^{m,l}$ and $E_S^{m,l}$, respectively.
All of them however are either product or compositions of functions.
By definition, for a PGF
\begin{align*}
    A(x,y) = \sum_{n = 0}^{n_\mathrm{max}} \sum_{m = 0}^{m_\mathrm{max}} a_{n,m} x^n y^m \;,
\end{align*}
we can extract the coefficient using
\begin{align*}
    a_{n,m} = \left . \frac{1}{n!m!}\frac{\partial^n\partial^m}{\partial x^n \partial y^m} A(x,y) \right |_{x,y = 0} \;.
\end{align*}
For product or composition of functions, this becomes impractical, as the number of terms for a partial derivative of degree $n$ requires $B_n$ terms, where $B_n$ is the Bell number.

Instead, let us use the characteristic function $\phi_{n,m}(u,v)$ for the random variables $n,m$, which can be written
\begin{align*}
    \phi_{n,m}(u,v) &= \left \langle e^{-2\pi i u n} e^{-2\pi i v m} \right \rangle \\
                    &= \sum_{n = 0}^{n_\mathrm{max}} \sum_{m = 0}^{m_\mathrm{max}} a_{n,m} e^{-2 \pi i u n } e^{- i 2 \pi v m} \\
                    &= A(e^{-i 2 \pi u}, e^{-i 2 \pi v}) \;.
\end{align*}
The coefficients $a_{n,m}$ can be recovered exactly using an inverse discrete Fourier transform,
\begin{align*}
    a_{n,m} &= \frac{1}{(n_\mathrm{max}+1)(m_\mathrm{max}+1)}\sum_{u = 0}^{n_\mathrm{max}} \sum_{v = 0}^{m_\mathrm{max}} \phi_{n,m}(u,v) e^{2 \pi i u n} e^{i 2 \pi v m} \;.
\end{align*}

\section{A-GAME: Additional results}

\begin{figure*}[t]
    \centering
    \includegraphics[width=\linewidth]{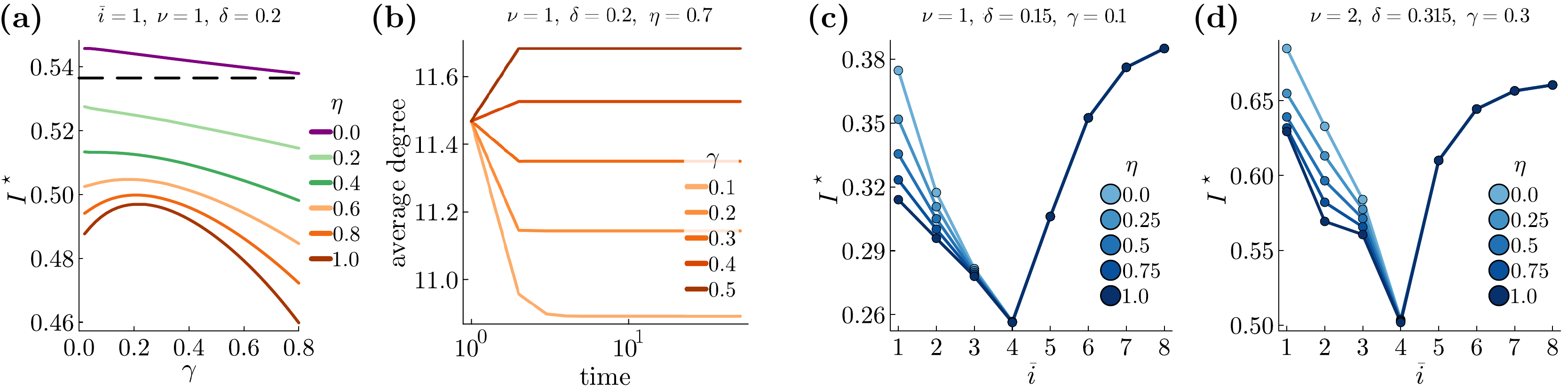}
    \caption{Further results for the adaptive hypergraphs in Fig.~3 of the main text.
    (a) Horizontal slices of the phase diagram in Fig.~3(a). The purple curve corresponds to detrimental rewiring, where the equilibrium fraction of infected nodes, $I^\star$, is larger than in absence of rewiring (dashed line); green curves represent increasingly beneficial rewiring; orange curves denote non-monotonic beneficial rewiring, where $I^\star$ first increases and then decreases.
    (b) Time evolution of the average degree for different rewiring rates $\gamma$ in the non-monotonic regime, $\eta = 0.7$.
    (c \& d) For slow rewiring, it is always optimal for agents to target the structure and minimize degree by setting $\bar{i} = 4 \approx \langle n\rangle$, no matter the rewiring accuracy.
    }
    \label{fig:adaptive}
\end{figure*}

We here provide additional results for the adaptive hypegraph model studied in the main text, which results are reported in Fig.~3 of the main text. Figure \ref{fig:adaptive}(a) helps distinguish the different regimes determined by the rewiring accuracy, $\eta$. In particular, we can appreciate the regime of high enough accuracy where an intermediate, least-optimal rewiring rate exists (orange curves). This emerges as an intermediary case where the two best strategies, i.e., targeting the dynamics (as for high enough $\gamma$) or the structure, are implemented in the least optimal way. Note that this is not the worst possible regime, as the local maximum still outperforms dynamics without any rewiring.

The fact that, in a region of low $\gamma$, the equilibrium fraction of infected nodes, $I^\star$, decreases by lowering $\gamma$ comes from the fact that a smaller $\gamma$ ensures a lower connectivity. Indeed, for very slow rewiring, at the typical rewiring time ($\gamma^{-1}$) many nodes are already infected (in fact, the slower the rewiring, the more the fraction of infected nodes overshoots initially) and the probability that a group includes at least $\nu$ infected nodes (and thus is infectious and avoided by susceptible nodes adopting $\bar{i}=\nu$) correlates already strongly with the size of the group. Consequently, especially for high rewiring accuracy, susceptible nodes will often escape large groups in favour of small groups, eventually leading to a more homogeneous group size distribution, hence to a lower average degree. When the rewiring rate is slightly increased, that correlation becomes weaker, in turn implying a less homogeneous group size distribution, thus a larger connectivity (see Fig.~\ref{fig:adaptive}(b)). However, rewiring is still too slow for the strategy $\bar{i}=\nu$ to work. Increasing the rewiring rate further, given the accuracy is high enough, the dynamics-targeting strategy $\bar{i}=\nu$ performs better and better, up to the point at which $I^\star$ decreases again with $\gamma$. In other words, the least optimal rewiring rate is too slow to avoid the dynamics but too fast to minimize the degree.

Finally, Figs.~\ref{fig:adaptive}(c \& d) show that minimizing connectivity, as implied by setting $\bar{i} = 4 \approx \langle n\rangle$, is always the best strategy when rewiring is slow, no matter how accurate the rewiring is.

